\documentclass[journal]{IEEEtran}
\usepackage{cite}
\usepackage{amsmath,amssymb,amsfonts}
\usepackage{graphicx}
\usepackage{textcomp}
\usepackage{xcolor}
\usepackage{todonotes}
\usepackage{tcolorbox}
\usepackage{url} 
\usepackage{stfloats}
\usepackage{booktabs}
\usepackage{algorithm}
\usepackage{algorithmic}
\usepackage{amsthm}
\usepackage{hyperref}
\usepackage[commandnameprefix=always]{changes}
\newtheorem{definition}{Definition}
\newtcolorbox{rqbox}[2][]{
  colframe=black!80!black,    
  colback=white,              
  coltitle=white,             
  colbacktitle=black,         
  boxrule=0.4mm,              
  title=#2,                   
  title filled=true,          
  #1
}
\definechangesauthor[name={Ruijun}, color=blue]{R}
\newcommand{\circled}[1]{\textcircled{\raisebox{0.1pt}{\scriptsize #1}}}

\ifCLASSOPTIONcompsoc
 \usepackage[caption=false,font=normalsize,labelfont=sf,textfont=sf]{subfig}
\else
 \usepackage[caption=false,font=footnotesize]{subfig}
\fi
\begin{document}

\title{CGP-Tuning: Structure-Aware Soft Prompt Tuning for Code Vulnerability Detection}

\author{Ruijun~Feng,
        Hammond~Pearce,
        Pietro~Liguori,
        Yulei~Sui
\thanks{This work was supported in part by the Australian Research Council Grants No. FT220100391 and No. DP250101396. H. Pearce is supported in part by a research gift from Intel Corporation and a gift from Google. (\textit{Corresponding author: Yulei Sui.})}
\thanks{R. Feng, H. Pearce and Y. Sui are with the School of Computer Science and Engineering, University of New South Wales (UNSW), Sydney, Australia (e-mails: \{ruijun.feng, hammond.pearce, y.sui\}@unsw.edu.au).}
\thanks{P. Liguori is with the Department of Electrical Engineering and Information Technology, University of Naples Federico II, Italy (e-mail: pietro.liguori@unina.it).}
}

\maketitle

\begin{abstract}
Large language models (LLMs) have been proposed as powerful tools for detecting software vulnerabilities, where task-specific fine-tuning is typically employed to provide vulnerability-specific knowledge to the LLMs. However, existing fine-tuning techniques often treat source code as plain text, losing the graph-based structural information inherent in code.

Graph-enhanced soft prompt tuning addresses this by translating the structural information into contextual cues that the LLM can understand. However, current methods are primarily designed for general graph-related tasks and focus more on adjacency information, they fall short in preserving the rich semantic information (e.g., control/data flow) within code graphs. They also fail to ensure computational efficiency while capturing graph-text interactions in their cross-modal alignment module.

This paper presents \textbf{CGP-Tuning}, a new code graph-enhanced, structure-aware soft prompt tuning method for vulnerability detection. CGP-Tuning introduces type-aware embeddings to capture the rich semantic information within code graphs, along with an efficient cross-modal alignment module that achieves linear computational costs while incorporating graph-text interactions. It is evaluated on the latest \textit{DiverseVul} dataset and three advanced open-source code LLMs, CodeLlama, CodeGemma, and Qwen2.5-Coder. Experimental results show that CGP-Tuning delivers model-agnostic improvements and maintains practical inference speed, surpassing the best graph-enhanced soft prompt tuning baseline by an average of four percentage points and outperforming non-tuned zero-shot prompting by 15 percentage points.
\end{abstract}

\begin{IEEEkeywords}
code large language model, soft prompt tuning, vulnerability detection, cross-modal alignment, graph neural network, multimodal learning.
\end{IEEEkeywords}

%
\IEEEpeerreviewmaketitle

\section{Introduction}
\IEEEPARstart{V}{ulnerability} detection has long been a complex challenge in software engineering. It involves identifying exploitable mistakes in given code snippets in advance of their misuse, which is essential for ensuring the reliability and integrity of software. One important approach to this is static analysis \cite{sui2016svf}. Modern static analysis approaches often convert source code into graph representations (i.e., code graphs) based on code structure and semantics, and then examine potential execution paths to identify vulnerabilities. While static analysis excels at identifying predefined exploitable patterns \cite{sui2012memoryleak, tomassi2021nullpointer}, its dependence on human-defined rules for analyzing code graphs limits its adaptability to new vulnerabilities.

Recent advances in code large language models (LLMs) \cite{roziere2024codellama, zhao2024codegemma, hui2024qwen25coder} have opened new possibilities for vulnerability detection. Leveraging the exceptional semantic understanding capabilities of code LLMs, simply feeding source code as plain text to these models has shown promising results in identifying simple vulnerabilities \cite{xin2024llm4vulnerability}. However, treating source code as plain text discards the rich graph-based structural information inherent in code, including its syntax, control flow, and data flow. Consequently, existing code LLMs lack a deep understanding of code structure and themselves lack mechanisms to natively process graph features.

One possible solution to this is to encode graphs as additional context and leverage the in-context learning capabilities of LLMs \cite{fatemi2024talk}. Lu et al. \cite{lu2024grace} introduced GRACE, which describes the code property graph in natural language and supplies it as additional context to LLMs. Their results demonstrate that integrating the graph-based structural information of source code can enhance LLM performance in vulnerability detection. However, GRACE relies heavily on the in-context learning capabilities of LLMs and is computationally expensive, with a high computational cost proportional to $(N + |V| + |E|)^2$ due to the quadratic complexity of self-attention \cite{vaswani2017attention}. Here, $N$ is the input sequence length, $V$ is the set of vertices (i.e., nodes), and $E$ is the set of edges.

A more efficient approach to integrating graph features with LLMs involves encoding these features into a fixed number of soft prompt embeddings, thereby reducing the context length required to describe them. This technique, known as graph-enhanced soft prompt tuning \cite{li2024graphpromptlearning}, often incorporates a cross-modal alignment module acting as a dynamic “interface" that translates graph-based structural information (i.e., graph features) into graph-enhanced soft prompt embeddings that align well with text-based input (i.e., text features). These embeddings serve as contextual cues that the LLMs can interpret, effectively bridging graph features with the semantic understanding of LLMs. Depending on the design of the cross-modal alignment module, these methods generally fall into two categories: projector-based methods \cite{chen2024llaga, tang2024graphgpt, ma2024xrec, he2024gretriever} and cross-attention-based methods \cite{tian2024gnp}.

In projector-based methods such as G-Retriever \cite{he2024gretriever}, the final node embeddings are first aggregated into a single graph embedding. A projector then maps this graph embedding into the latent space of the LLM as a graph-enhanced soft prompt embedding. This cross-modal alignment process is computationally efficient, scaling linearly with the number of nodes $|V|$. However, it aligns only the graph features into the LLM's latent space to make them interpretable by the model, without modeling their interactions with text features.

In contrast, cross-attention-based methods explicitly model the interactions between graph and text features through cross-attention mechanisms. For instance, in graph neural prompting (GNP) \cite{tian2024gnp}, final node embeddings first undergo self-attention to account for node significance. The output then serves as queries to perform cross-attention with the text features, producing aligned node embeddings that are aggregated and projected into a graph-enhanced soft prompt embedding. While this cross-modal alignment process incurs a higher multiplicative computational cost proportional to $|V|^2 + |V| \times N + |V|$, it more effectively captures the correlations between graph and text features.

Both categories have demonstrated strong performance on general graph-related tasks (e.g., recommendation systems and social network analysis) \cite{chen2024llaga, tang2024graphgpt, ma2024xrec, he2024gretriever, tian2024gnp}. However, they face two major limitations when applied to vulnerability detection:

\begin{itemize}
    \item \textbf{Limited ability to effectively capture semantic information within code graphs}: These methods are primarily designed for general graph-related tasks, with a strong emphasis on adjacency information. When applied directly to vulnerability detection, they fail to fully exploit the rich semantic information inherent in code graphs.
    
    \item \textbf{Fail to ensure computational efficiency while considering graph-text interactions}: These methods either ignore graph-text interactions, as in projector-based methods, or incur high computational costs when such interactions are considered, as in cross-attention-based methods.
\end{itemize}

To address these limitations, this paper proposes a code graph-enhanced, structure-aware soft prompt tuning method, \textbf{CGP-Tuning}, which is designed to enhance code structure understanding in code LLMs for more effective vulnerability detection. The main contributions of this work are as follows:

\begin{enumerate}
    \item Code property graph~\cite{fabian2014cpg} is used to represent the graph-based structural information of source code. It is a superset that combines multiple code representations, including abstract syntax tree, control flow graph, and program dependence graph.
    
    \item Two trainable, type-aware embeddings are proposed to enable the model to capture rich semantic information from node and edge types in the code property graph.
    
    \item An efficient cross-modal alignment module is proposed, with a linear computational cost scaling with $|V| + N$, while considering graph-text interactions.
    
    \item A comprehensive and in-depth analysis of CGP-Tuning's effectiveness is conducted, including a comparison with other state-of-the-art methods, an assessment of each component, an evaluation of its performance on long source code, and an analysis of its total inference time relative to other approaches.
        
    \item The implementations of CGP-Tuning are made available at: \href{https://github.com/ruijunfeng/CGP-tuning}{https://github.com/ruijunfeng/CGP-tuning}.
\end{enumerate}

\section{Background and Motivation}
This section introduces the background knowledge of graph-enhanced soft prompt tuning methods and outlines the motivation behind the proposed CGP-Tuning.

\subsection{Graph Neural Network}
Graph neural networks \cite{kipf2017gcn, velickovic2018gat} are specialized neural networks designed for processing graph-structured data, where information is organized as a set of vertices (i.e., nodes) and edges (i.e., connections between nodes). Each node is initialized with a node embedding that is iteratively updated through \textbf{message passing}, where each node aggregates information from its neighbouring nodes (i.e., adjacency information). This process enables the final node embeddings to capture both each node's intrinsic features and its local topology, thereby reflecting the graph-based structural information of source code suitable for various software engineering tasks.

For node-level tasks like fault localization \cite{gou2024faultlocalization}, final node embeddings are individually leveraged for statement-level fault detection. For graph-level tasks like vulnerability detection \cite{cheng2021deepwukong, xiao2024enhancedgraph, qiu2024multiplegraph, chakraborty2022reveal}, final node embeddings are pooled into a single, comprehensive graph-level embedding for function-level vulnerability detection. Many graph-enhanced soft prompt tuning methods rely on graph neural networks to extract different node-level and graph-level features for LLMs \cite{tang2024graphgpt, ma2024xrec, he2024gretriever, tian2024gnp}.
\begin{figure}[h!]
    \centering
    \includegraphics[width=0.7\linewidth]{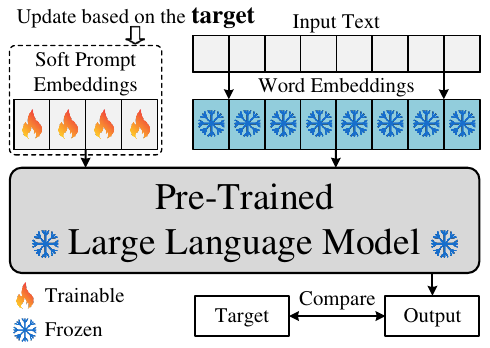}
    \caption{Example of soft prompt tuning (i.e., prompt tuning).}
    \label{fig:soft_prompt_tuning}
\end{figure}

\subsection{Code Large Language Model}
Unlike earlier code language models such as CodeBERT \cite{feng2020codebert} and CodeT5 \cite{wang2021codet5}, which contain only millions of parameters. Modern code LLMs like CodeLlama \cite{roziere2024codellama}, CodeGemma \cite{zhao2024codegemma}, and Qwen2.5-Coder \cite{hui2024qwen25coder} have been scaled up to billions of parameters. This dramatic increase in scale allows modern code LLMs to more effectively learn programming patterns and semantic relationships through pre-training on source code and related documentation \cite{roziere2024codellama, zhao2024codegemma, hui2024qwen25coder}.

After task-specific fine-tuning, these code LLMs demonstrate state-of-the-art performance on various software engineering tasks such as code generation \cite{lam2024llmrpg}, where they assist developers by automating the creation of syntactically and semantically correct code, and code review \cite{yu2024codereview, nashaat2024codementor}, where they analyze existing code to identify potential errors.

However, this enhanced performance comes at the cost of significantly greater resource requirements. For instance, simply storing an LLM with seven billion parameters requires 28 gigabytes of memory under Float32 precision, which exceeds the capacity of most consumer-level hardware. This limitation makes traditional full-parameter fine-tuning \cite{feng2020codebert, wang2021codet5, guo2021graphcodebert, thapa2022vulcodebert}, which updates all parameters in the model, impractical for many use cases.

To address these challenges, parameter-efficient fine-tuning techniques \cite{ding2023peft}, such as soft prompt tuning \cite{liu2022ptuning, lester2021prompttuning, li2021prefixtuning}, have emerged as more scalable and accessible alternatives. These methods adjust only a small set of additional parameters, enabling the adaptation of pre-trained code LLMs to new tasks while significantly reducing resource demands.
\begin{figure*}[t]
\centering

\subfloat[Projector-based cross-modal alignment module.]{
    \includegraphics[width=0.23\textwidth]{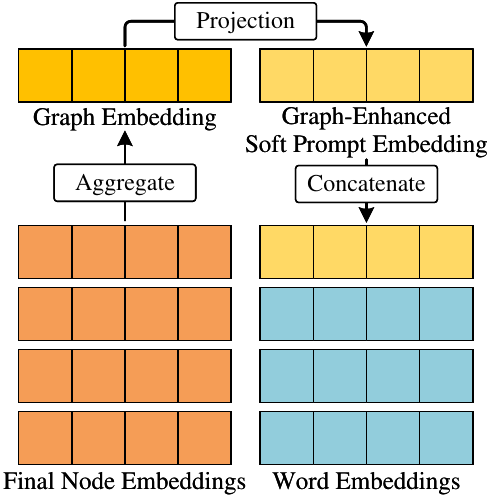}
    \label{fig:projector_based}
}
\hfill
\subfloat[Cross-attention-based cross-modal alignment module.]{
    \includegraphics[width=0.74\textwidth]{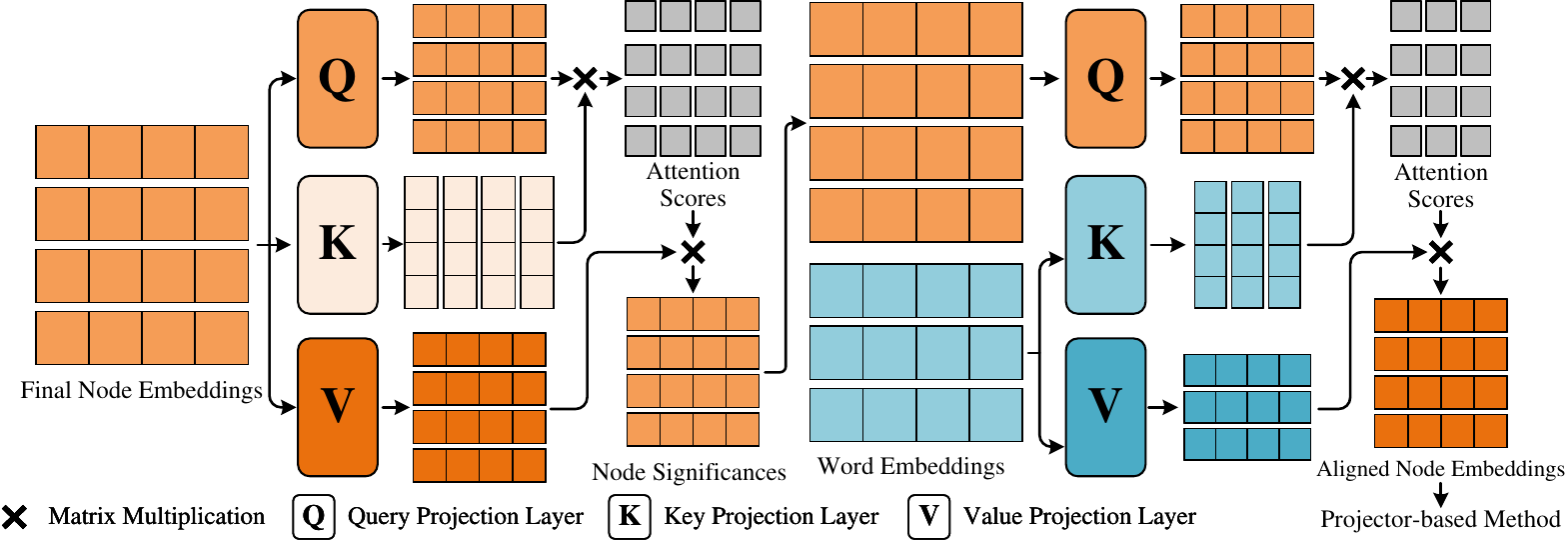}
    \label{fig:cross_attention_based}
}
\vspace{1em} 

\caption{Comparison of different cross-modal alignment modules.}
\label{fig:cross_modal_alignment_module}
\end{figure*}

\begin{figure*}[t]
    \centering
    \includegraphics[width=\textwidth]{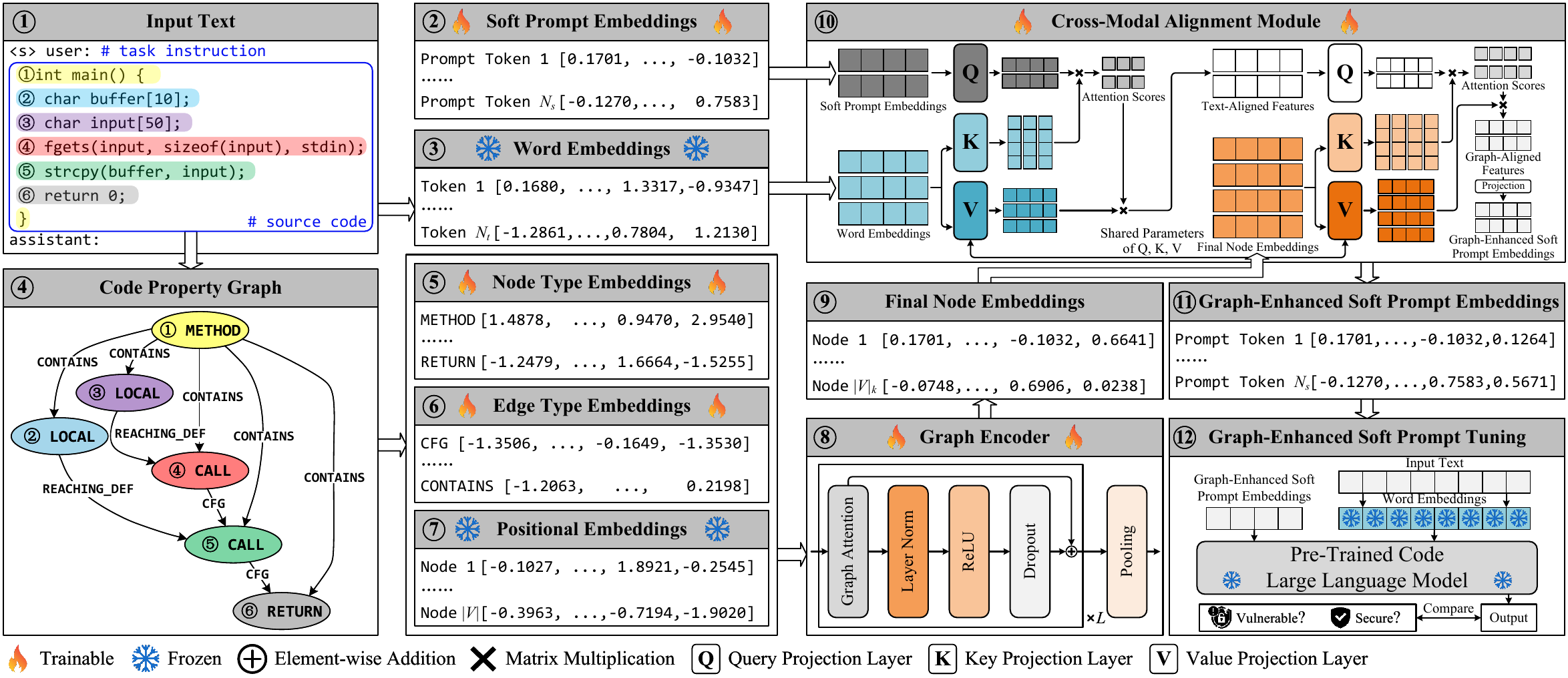}
    \caption{Overview of CGP-Tuning.}
    \label{fig:cgp_tuning}
\end{figure*}

\subsection{Soft Prompt Tuning}
Soft prompt tuning \cite{liu2022ptuning, li2021prefixtuning, lester2021prompttuning} is a parameter-efficient fine-tuning method that uses soft prompt embeddings at different locations to adapt pre-trained LLMs. For example, Figure \ref{fig:soft_prompt_tuning} shows the process of prompt tuning \cite{lester2021prompttuning}. It prepend soft prompt embeddings to the word embeddings of the input text (e.g., task instruction, data sample) as an enhancement. During fine-tuning, only the trainable parameters used to generate soft prompt embeddings are updated while the pre-trained LLMs and word embeddings remain frozen \cite{liu2022ptuning, li2021prefixtuning, lester2021prompttuning, wang2023ptuning4code}.

Soft prompt tuning has proven effective across various software engineering tasks, such as defect detection and code translation \cite{wang2023ptuning4code}, often achieving performance comparable to full-parameter fine-tuning \cite{liu2022ptuning, wang2023ptuning4code}. Although fine-tuning can provide vulnerability-specific knowledge to pre-trained code LLMs, they still lack a deep understanding of code structure. Since modern code LLMs, including those fine-tuned with soft prompt tuning \cite{liu2022ptuning, li2021prefixtuning, lester2021prompttuning} or other fine-tuning techniques \cite{feng2020codebert, wang2021codet5, hu2022lora}, cannot alter the input modality of the LLMs. They strictly follow the original text-based workflow and cannot directly leverage the rich graph-based structural information.

\subsection{Graph-Enhanced Soft Prompt Tuning} \label{sec:graph_enhanced_soft_prompt}
Graph-enhanced soft prompt tuning \cite{li2024graphpromptlearning} addresses this limitation by incorporating structural information from graphs. This is achieved by encoding graph features using either graph neural networks \cite{tang2024graphgpt, ma2024xrec, he2024gretriever, tian2024gnp} or flattened graph sequences \cite{chen2024llaga}, followed by a cross-modal alignment module to align graph features with the text-based input. Depending on the alignment strategy, existing methods can be classified as projector-based \cite{chen2024llaga, tang2024graphgpt, ma2024xrec, he2024gretriever} or cross-attention-based \cite{tian2024gnp}.

Figure \ref{fig:projector_based} illustrates the alignment process of projector-based methods like G-Retriever \cite{he2024gretriever}, which is designed for graph-level tasks like vulnerability detection. In this approach, a graph neural network first computes $|V|$ final node embeddings (i.e., graph features) and aggregates them into a single graph embedding. This embedding is then mapped by a projector into the LLM's latent space as a graph-enhanced soft prompt embedding, which is concatenated with word embeddings (i.e., text features) for enhancement. It can be observed that this cross-modal alignment process incurs a linear computational cost proportional to $|V|$, but it fails to account for interactions between graph and text features.

Cross-attention-based methods, exemplified by GNP \cite{tian2024gnp}, address the aforementioned limitation by explicitly modeling interactions between graph and text features. As depicted in Figure \ref{fig:cross_attention_based}, GNP first applies self-attention over the $|V|$ final  node embeddings to capture their relative importance. The resulting node significance matrices is then used as queries, with word embeddings serving as key-value pairs for cross-attention. The output aligned node embeddings are aggregated and projected in the same manner as projector-based methods (Figure \ref{fig:projector_based}). This cross-modal alignment process enables GNP to effectively capture graph-text interactions but incurs a multiplicative computational cost proportional to $|V|^2 + |V| \times N + |V|$. Hence, when processing long source code with complex code graphs, it requires significantly more memory to store attention scores of size $|V|^2$ and $|V| \times N$.

The proposed CGP-Tuning aims to account for graph-text interactions while maintaining computational efficiency by decomposing the computations required for cross-attention and projection, the details of which can be found in Section \ref{sec:cross_modal_alignment}.

\subsection{Example}
To illustrate the differences between CGP-Tuning and several previous methods for integrating graph-based structural information into code LLMs, Figure \ref{fig:cgp_tuning} presents an overview of CGP-Tuning (see Section \ref{sec:approach} for details). The key distinctions of CGP-Tuning compared to G-Retriever and GNP lie in the cross-modal alignment modules (step \circled{10}), as described in Section \ref{sec:graph_enhanced_soft_prompt}, and the omission of node and edge type information (steps \circled{5} and \circled{6}).

Specifically, GNP uses the cross-modal alignment module in Figure \ref{fig:cross_attention_based} to replace step \circled{10} and excludes step \circled{2}, as it relies on a heavier self-attention to capture node significance and uses cross-attention to model graph-text interactions. G-Retriever employs a lightweight cross-modal alignment module in Figure \ref{fig:projector_based} to replace step \circled{10} and excludes steps \circled{2} and \circled{3}, as it does not model graph-text interactions during projection. Moreover, both G-Retriever and GNP exclude steps \circled{5} and \circled{6}, as they are designed for general graph-related tasks that focus on adjacency information rather than the semantic information in code graphs. CGP-Tuning includes all steps in Figure \ref{fig:cgp_tuning}, making it the most comprehensive approach, fully leveraging the semantic information in code graphs while capturing graph-text interactions at a lower cost than GNP.

\begin{figure}[t]
\centering

\subfloat[Input-output example (general).]{
    \includegraphics[width=0.75\columnwidth]{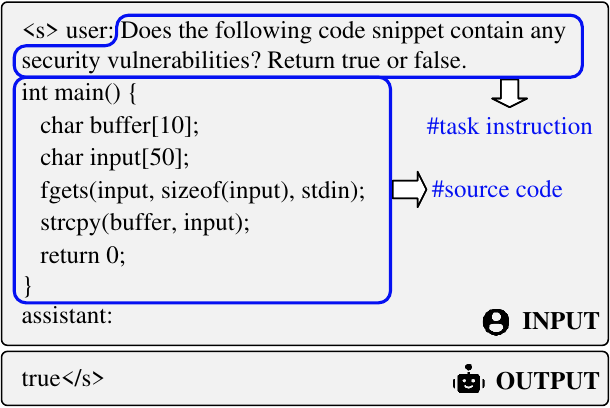}
    \label{fig:input_output_example}
}

\subfloat[Input-output example (GRACE).]{
    \includegraphics[width=0.75\columnwidth]{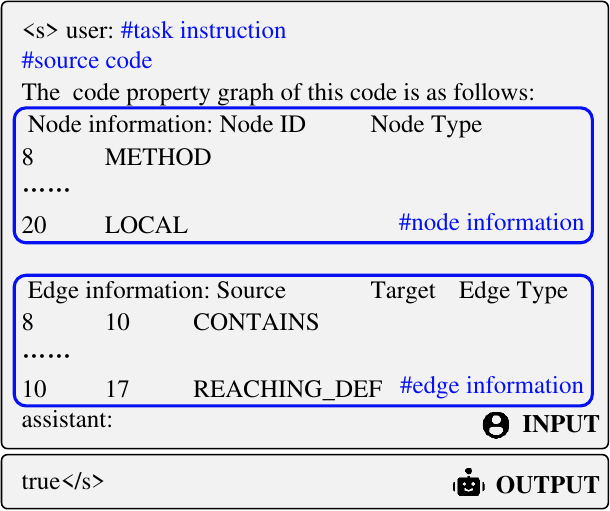}
    \label{fig:input_output_example_grace}
}

\caption{Input-output examples of different methods.}
\label{fig:input_output_example_of_different_method}
\end{figure}

Figure \ref{fig:input_output_example_of_different_method} shows two input-output examples for further comparison. For G-Retriever, GNP, and CGP-Tuning (Figure \ref{fig:input_output_example}), the code LLMs receive a task instruction and source code wrapped in a chat template, then return their predictions. These methods generate virtual, graph-enhanced soft prompt embeddings to incorporate structural information as described above, so no readable graph-enhanced prompt content is visible in the example. In contrast, GRACE \cite{lu2024grace} represents the code property graph in natural language, providing code LLMs with explicit contextual information about nodes and edges (Figure \ref{fig:input_output_example_grace}).

\section{Approach} \label{sec:approach}
\subsection{Definitions}
Given a dataset $\mathcal{D}$ consisting of multiple input-output pairs ($X$, $Y$), the input text $X$ is source code wrapped in a chat template. This input text is first tokenized into a series of $N_t$ input text tokens, which are then transformed into word embeddings $X_t \in \mathbb{R}^{N_t \times d_\text{lm}}$ based on vocabulary $\mathcal{W}$, where $d_\text{lm}$ is the dimensionality of the word embeddings. The corresponding $Y$ represents the ground truth. A set of soft prompt embeddings, denoted as $X_s \in \mathbb{R}^{N_s \times d_{\text{lm}}}$, is initialized with trainable parameters $\theta_s$ and shares the same dimensionality as the word embeddings (Figure \ref{fig:cgp_tuning}, steps \circled{1} to \circled{3}).

The code property graph of the source code is represented as $G = (V, E, \tau_V, \tau_E)$, where $V$ is the set of nodes and $E$ is the set of edges. Each node $v_i \in V$ corresponds to a set of code statements, and each edge $e_{ij} \in E$ describes the relationships between nodes $v_i$ and $v_j$ (e.g., syntactic structure, control flow, or program dependencies). The function $\tau_V: V \to T_V$ assigns a node type from a set of node types $T_V$ to each node $v_i$ in $V$, while the function $\tau_E: E \to T_E$ assigns an edge type from a set of edge types $T_E$ to each edge $e_{ij}$ in $E$. Here, $\tau_V(v_i)$ and $\tau_E(e_{ij})$ specify the node type of node $v_i$ and the edge type of edge $e_{ij}$, respectively (Figure \ref{fig:cgp_tuning}, step \circled{4}).

In traditional soft prompt tuning (i.e., prompt tuning) \cite{lester2021prompttuning}, the objective is to learn the optimal parameters $\theta_s$ as soft prompt embeddings $X_s$, such that the expected loss over the dataset, denoted $\mathbb{E}_{(X,Y) \sim \mathcal{D}}$, is minimized. Here, the expectation is computed using the loss function $\mathcal{L}$, averaged over all input-output pairs ($X$, $Y$) in $\mathcal{D}$.

\begin{definition}[\textbf{Traditional Soft Prompt Tuning}]
\[
\min_{\theta_s} \; \mathbb{E}_{(X,Y) \sim \mathcal{D}} \left[ \mathcal{L}\big( \textnormal{CodeLLM}([X_s; X_t]), Y \big) \right]
\]
\end{definition}

Unlike traditional soft prompt tuning, which employs $\theta_s$ solely to find optimal $X_s$, graph-enhanced soft prompt tuning typically aims to learn a set of parameters $\theta = \{\theta_g, \theta_c\}$, where $\theta_g$ corresponds to the parameters of graph encoder and $\theta_c$ corresponds to the parameters of cross-modal alignment module. The graph-enhanced soft prompt embedding $Z \in \mathbb{R}^{1 \times d_{\text{lm}}}$ is typically obtained via $Z = f_{\theta_g, \theta_c}(G)$ in the projector-based method \cite{he2024gretriever}, whereas cross-attention-based method \cite{tian2024gnp} derives it by $Z = f_{\theta_g, \theta_c}(G, X_t)$.

\begin{definition}[\textbf{Graph-Enhanced Soft Prompt Tuning}]
\[
\min_{\theta} \; \mathbb{E}_{(X,Y) \sim \mathcal{D}} \left[ \mathcal{L}\big( \textnormal{CodeLLM}([Z; X_t]), Y \big) \right]
\]
\end{definition}

Moving beyond these approaches, CGP-Tuning aims to learn the optimal set of parameters $\theta = \{\theta_s, \theta_t, \theta_g, \theta_c\}$, where $\theta_t$ denotes the parameters for type-aware embeddings. The method captures type-specific information within the code property graph via $\theta_t$, while leveraging features from both graph $G$ and text $X_t$ to generate more informative graph-enhanced soft prompt embeddings $Z \in \mathbb{R}^{N_s \times d_{\text{lm}}}$ through $Z = f_{\theta_t, \theta_g, \theta_c}(X_s, G, X_t)$ with low computational costs.

\subsection{Graph Representation}
To capture the semantic information of node types and edge types within the code property graph, this paper proposes type-aware embeddings that map node types and edge types into their respective feature spaces, as follows:
\begin{align} 
     X_{T_V} &= \{x_{t_v} \mid t_v \in T_V\} \in \mathbb{R}^{|T_V| \times d_{t_v}} \label{eq:node_type_embedding}\\
     X_{T_E} &= \{x_{t_e} \mid t_e \in T_E\} \in \mathbb{R}^{|T_E| \times d_{t_e}} \label{eq:edge_type_embedding}
\end{align}
where $d_{t_v}$ and $d_{t_e}$ represent the dimensions of the node and edge type embeddings, respectively. For example, given a node $v_i$ (edge $e_{ij}$), its corresponding node (edge) type embedding is represented as $x_{\tau_V(v_i)} \in \mathbb{R}^{1 \times d_{t_v}}$ ($x_{\tau_E(e_{ij})} \in \mathbb{R}^{1 \times d_{t_e}}$).

In these type-aware embedding layers (Figure \ref{fig:cgp_tuning}, steps \circled{5} and \circled{6}), the semantic relationships between node types and edge types are encoded separately into their respective embeddings. These type-aware embeddings allow the graph neural network to distinguish between different types of nodes and edges during message passing and feature aggregation. By mapping node and edge types into trainable dense embeddings, the model can effectively capture type-specific semantic information, which is crucial for understanding the structural and functional dependencies within the code property graph.

Then, for each node $v_i$, a sinusoidal positional embedding \cite{vaswani2017attention} $pe_i$ is computed, and the set of these embeddings is denoted as $PE = \{pe_i \mid i = 1, \dots, |V|\} \in \mathbb{R}^{|V| \times d_{pe}}$ (Figure \ref{fig:cgp_tuning}, step \circled{7}). Each $pe_i$ is initialized as follows:
\begin{align}
    pe_{i, 2j} &= \sin\left(i \cdot 10000^{-2j / d_{pe}}\right) \label{eq:sin_positional_embedding} \\
    pe_{i, 2j+1} &= \cos\left(i \cdot 10000^{-2j / d_{pe}}\right) \label{eq:cos_positional_embedding}
\end{align}
where $j$ denotes the $j$-th dimension of $pe_i$. The dimensions of the node type embeddings and positional embeddings are set to match those of the word embeddings used in the code LLM for consistency, i.e., $d_{t_v} = d_{pe} = d_\text{lm}$.

The initial node embeddings $H_1 = \{h_{1, i} \mid i = 1, \dots, |V|\} \in \mathbb{R}^{|V| \times d_\text{lm}}$ are obtained by adding the node type embedding $x_{\tau_V(v_i)}$ and positional embedding $pe_i$ as follows:
\begin{equation} 
    h_{1, i} = x_{\tau_V(v_i)} + pe_i \label{eq:node_embed} 
\end{equation}

While node type embeddings $x_{\tau_V(v_i)}$ may be shared among multiple nodes of the same type, the positional embeddings $pe_i$ are inherently distinct, as the index $i$ of each node is unique. This guarantees the uniqueness of the resulting node embeddings and makes them well-suited for length extrapolation.

These initial node embeddings, along with the edge type embeddings, serve as input to the subsequent graph encoder to obtain the final node embeddings as graph features.

\subsection{Graph Encoder} \label{sec:graph_encoder}
Step \circled{8} in Figure \ref{fig:cgp_tuning} shows the overall structure of the graph encoder. Specifically, this graph encoder employs a basic block composed of the graph attention network (GAT) \cite{velickovic2018gat}, layer normalization (LN) \cite{ba2016layernormalization}, the ReLU activation function \cite{agarap2019relu}, and a dropout layer \cite{srivastava2014dropout}, defined as follows: 
\begin{equation} 
    \tilde{H}^{(l+1)} = \text{Dropout}(\text{ReLU}(\text{LN}(\text{GAT}(H^{(l)}, X_{\tau_E(E)})))) \label{eq:gnn_basic_block}
\end{equation}

To alleviate the over-smoothing problem \cite{wu2023oversmoothing}, a residual connection \cite{he2016residual} is employed within each basic block, defined as follows: 
\begin{equation}
    H^{(l+1)} = \tilde{H}^{(l+1)} + H^{(l)}, \quad l = 1, \dots, L \label{eq:gnn_residual_link}
\end{equation}
where $L$ is the total number of basic blocks.

The initial input to these basic blocks, $H^{(1)}$, is $H_1$, and the output after stacking $L$ basic blocks is denoted as $H_2 \in \mathbb{R}^{|V| \times d_\text{lm}}$, where $H_2$ corresponds to $H^{(L)}$. These basic blocks are stacked to iteratively extract graph features.

Finally, a pooling layer \cite{ranjan2020asap} is applied to produce the final node embeddings (Figure \ref{fig:cgp_tuning}, step \circled{9}):
\begin{equation}
    H_3 = \text{POOL}(H_2) \label{eq:graph_pool}
\end{equation}
where $H_3 \in \mathbb{R}^{|V|_k \times d_\text{lm}}$ and $|V|_k = \min(k, |V|)$. This pooling layer selects up to $k$ of the most important node groups in the graph by identifying and scoring local regions based on their relevance to vulnerability detection. The final output, $H_3$, retains only the top-scoring groups while preserving structural information, resulting in a compact and informative representation of the code property graph that facilitates effective integration with subsequent components.

\subsection{Cross-Modal Alignment Module} \label{sec:cross_modal_alignment}
To align the extracted graph features $H_3 \in \mathbb{R}^{|V|_k \times d_\text{lm}}$ with the text features $X_t \in \mathbb{R}^{N_t \times d_\text{lm}}$, this paper proposes an efficient cross-modal alignment module (Figure \ref{fig:cgp_tuning}, step \circled{10}).

As described in Section \ref{sec:graph_enhanced_soft_prompt}, traditional cross-attention-based method \cite{tian2024gnp} incurs a multiplicative computational cost that scales with $|V|^2 + |V| \times N_t + |V|$ due to its inefficient computational process in cross-attention and projection. The proposed method improves this by introducing soft prompt embeddings $X_s$ as a global context (Figure \ref{fig:cgp_tuning}, step \circled{2}), thereby decomposing the required computation. The following equations define how the method considers graph-text interactions without direct cross-attention between graph and text features:
\begin{align}
    H_4 &= \text{MultiHeadAttention}({X_s}{W_Q}, {X_t}{W_K}, {X_t}{W_V})  \label{eq:text_feature}\\
    H_5 &= \text{MultiHeadAttention}({H_4}{W_Q}, {H_3}{W_K}, {H_3}{W_V})  \label{eq:graph_feature}
\end{align}

It starts with three shared weight matrices, $W_Q$, $W_K$, and $W_V$, each with dimensions $\mathbb{R}^{d_\text{lm} \times d_\text{lm}}$. These matrices transform the soft prompt embeddings, graph features, and text features into the same subspaces for attention computation. 

In equation \ref{eq:text_feature}, $X_s$ acts as the queries while the text features $X_t$ (Figure \ref{fig:cgp_tuning}, step \circled{3}) serve as the key-value pairs to perform attention calculation, producing the text-aligned features $H_4 \in \mathbb{R}^{N_s \times d_\text{lm}}$ that capture dependencies with each word in the textual content. This process incurs a computational cost proportional to $N_t \times N_s$. 

Subsequently, in equation \ref{eq:graph_feature}, $H_4$ is used as the queries to perform attention calculation with the graph features $H_3$ (Figure \ref{fig:cgp_tuning}, step \circled{9}), yielding the graph-aligned features $H_5 \in \mathbb{R}^{N_s \times d_\text{lm}}$ that model the relevance to each node in the graph, with a computational cost proportional to $|V|_k \times N_s$. These two steps successfully capture the graph-text interactions.

Finally, a feed-forward network acts as a projector, mapping the graph-aligned features into the latent space of the code LLM as the final graph-enhanced soft prompt embeddings $Z \in \mathbb{R}^{N_s \times d_\text{lm}}$ (Figure \ref{fig:cgp_tuning}, step \circled{11}):
\begin{equation}
    Z = \text{FeedForwardNetwork}(H_5) \label{eq:ffn}
\end{equation}

The computational cost of the entire process is proportional to $N_t \times N_s  + |V|_k \times N_s + N_s$. Since the number of soft prompt embeddings, $N_s$, remains constant throughout the process, only the number of input text tokens and the number of nodes vary with the input. This results in a linear computational cost proportional solely to $|V|_k + N_t$. The corresponding graph-enhanced soft prompt embeddings, $Z$, now incorporate the graph-based structural information of source code.

\subsection{Graph-Enhanced Soft Prompt Tuning}
Given the graph-enhanced soft prompt embeddings $Z$ and the word embeddings $X_t$, these embeddings are concatenated and fed into the pre-trained code LLM, producing the output logits $\hat{Y} \in \mathbb{R}^{(N_s + N_t) \times {|\mathcal{W}|}}$, as defined below:
\begin{equation}
    \hat{Y} = \textnormal{CodeLLM}([Z; X_t]) \label{eq:codellm}
\end{equation}

The output logits $\hat{Y}$ are then used to compute the cross-entropy loss $\mathcal{L}$ with the ground truths $Y$, as defined below:
\begin{equation} 
    \mathcal{L}(Y,\hat{Y}) = -\frac{1}{|T|} \sum_{i \in T} \sum_{j=1}^{{|\mathcal{W}|}} y_{ij} \log \hat{y}_{ij} \label{eq:cross_entropy}
\end{equation}
where $T$ is the set of indices excluding the soft prompt embeddings and the word embeddings of input text tokens for supervised fine-tuning and $|\mathcal{W}|$ is the vocabulary size.

Only the parameters of the soft prompt embeddings ($\theta_s$), type-aware embeddings ($\theta_t$), the graph encoder ($\theta_g$), and the cross-modal alignment module ($\theta_c$) are updated via backpropagation. This process enhances the code LLM's understanding of code structure and improves its ability to detect vulnerabilities, thereby achieving a code graph-enhanced, structure-aware soft prompt tuning (Figure \ref{fig:cgp_tuning}, step \circled{12}).

\section{Experiment Design}
This section provides details of the experimental design, including research questions, dataset descriptions, baselines, evaluation metrics, and implementation details.

\subsection{Research Questions}
To evaluate the effectiveness of the proposed CGP-Tuning, four research questions are posed as follows:

\begin{itemize}
    \item \textbf{RQ1}: Does CGP-Tuning outperform both traditional methods and state-of-the-art graph-enhanced soft prompt tuning methods in vulnerability detection?
    
    \item \textbf{RQ2}: Does incorporating the cross-modal alignment module, type-aware embeddings, and positional embeddings effective in improving the performance of CGP-Tuning?
    
    \item \textbf{RQ3}: Does CGP-Tuning reduce computational costs in its cross-modal alignment module at the expense of performance when processing long source code?

    \item \textbf{RQ4}: Does CGP-Tuning's computational efficiency consistent with the theoretical expectations?
\end{itemize}

\subsection{Dataset and Code Property Graph}
This paper utilizes the latest vulnerability detection dataset, DiverseVul \cite{chen2023diversevul}, which is derived from 809 projects and includes 150 distinct Common Weakness Enumerations. The dataset is split into training, validation, and testing subsets using a 70\%-15\%-15\% ratio. The entire dataset comprises over 330,000 samples, of which more than 94\% are non-vulnerable, resulting in a highly imbalanced class distribution. To address this imbalance, under-sampling is applied to each subset by removing non-vulnerable samples \cite{sultana2024codevulnerabilitydetectioncomparative}.

The code property graph of each data sample is generated using Joern v2.0.448\footnote{\url{https://github.com/joernio/joern}}, a tool capable of parsing various programming languages into code property graphs\footnote{\url{https://cpg.joern.io/}}. To prevent out-of-memory errors during training caused by large code graphs, samples with over 300,000 nodes and 30,000 edges are excluded from the training and validation sets. Tables \ref{tab:node_type_description} and \ref{tab:edge_type_description} summarize the node and edge types in the code property graphs of all samples after dataset processing.

\begin{table*}[ht]
\centering
\caption{Node Type Summary in Code Property Graph}
\label{tab:node_type_description}
\begin{tabular}{@{}l|l|c@{}}
\toprule
\textbf{Node Type} & \textbf{Description} & \textbf{Total} \\ 
\midrule
CONTROL\_STRUCTURE & Represents control constructs like loops (for, while) and conditionals (if, else, switch). & 436,927 \\
METHOD & Represents a function or method definition. & 69,601 \\
METHOD\_REF & References to methods for use in callbacks or function pointers. & 143 \\
METHOD\_RETURN & Return value of the method, usually a placeholder ``RET". & 69,601 \\
METHOD\_PARAMETER\_IN & Input parameters of a method. & 81,677 \\
METHOD\_PARAMETER\_OUT & Output parameters modified and returned by a method. & 81,677 \\
IDENTIFIER & General identifiers, including variable names. & 2,274,670 \\
LOCAL & Declares local variables within a function or block scope. & 170,102 \\
FIELD\_IDENTIFIER & Refers to fields or members of a structure or class. & 610,690 \\
CALL & Represents function or method invocations. & 2,570,335 \\
BLOCK & Groups sequences of statements, typically enclosed in braces. & 448,979 \\
RETURN & Indicates return statements within methods. & 85,211 \\
LITERAL & Represents constant values such as integers and strings. & 547,622 \\
TYPE\_DECL & Declares types like classes, structures, or union. & 68 \\
MODIFIER & Specifies modifiers like public, private, static, etc., for declarations. & 11,863 \\
JUMP\_TARGET & Marks targets for jump statements like goto or labeled statements. & 80,201 \\
UNKNOWN & Catches abstract syntax tree nodes that could not be categorized. & 102,285 \\
\bottomrule
\end{tabular}
\end{table*}

\begin{table*}[ht]
\centering
\caption{Edge Type Summary in Code Property Graph}
\label{tab:edge_type_description}
\begin{tabular}{@{}l|l|c@{}}
\toprule
\textbf{Edge Type} & \textbf{Description} & \textbf{Total} \\ 
\midrule
AST & Represents parent-to-child relationships in the abstract syntax tree. & 7,614,495 \\
CONTAINS & Connects a node to the method or structure containing it. & 7,188,846 \\
CFG & Represents the control flow between source nodes and target nodes. & 6,871,046 \\
DOMINATE & Represents immediate dominance from the source node to the target node in the control flow graph. & 6,388,200 \\
POST\_DOMINATE & Represents immediate post-dominance from the source node to the target node in the control flow graph. & 6,419,025 \\
CDG & Indicates that the target node is control dependent on the source node. & 6,503,863 \\
REACHING\_DEF & Data dependence edges showing variable definitions reaching other nodes without reassignment. & 21,281,599 \\
CALL & Links call sites (``CALL" node) to the methods they invoke. & 876 \\
ARGUMENT & Connects call sites (``CALL" node) to their arguments (``EXPRESSION" node). & 4,976,727 \\
RECEIVER & Links call sites to the receiver object. & 18,319 \\
REF & Indicates a reference from an identifier to the entity it denotes. & 1,749,523 \\
PARAMETER\_LINK & Connects input parameters to their corresponding output parameters in methods. & 81,677 \\
CONDITION & Connects control structures to their condition expressions. & 289,040 \\
\bottomrule
\end{tabular}
\end{table*}

\subsection{Baselines and Evaluation Metrics}
The evaluation uses accuracy, macro-average precision, recall, and F1-score as metrics. The seven-billion-parameter (7B) base versions of CodeLlama\footnote{\url{https://huggingface.co/meta-llama/CodeLlama-7b-hf}} \cite{roziere2024codellama}, CodeGemma\footnote{\url{https://huggingface.co/google/codegemma-7b}} \cite{zhao2024codegemma}, and Qwen2.5-Coder\footnote{\url{https://huggingface.co/Qwen/Qwen2.5-Coder-7B}} \cite{hui2024qwen25coder} serve as the evaluated code LLMs. These 7B models were chosen to balance computational feasibility with strong performance. Notably, the Qwen2.5-Coder 7B version represents the current state of the art, demonstrating superior capabilities compared to some outdated, larger models, such as the WizardCoder 15B version. It also outperforms the larger and more recent DeepSeek-Coder-V2 16B version across more than 20 distinct benchmarks \cite{hui2024qwen25coder}. To obtain structured output from code LLMs, a constrained decoding strategy similar to Jsonformer\footnote{\url{https://github.com/1rgs/jsonformer}} is adopted, wherein the classification result is determined by comparing the logits corresponding to the “true" and “false" tokens.

\begin{figure}[t]
\centering

\subfloat[Chat template.]{
    \includegraphics[width=0.75\columnwidth]{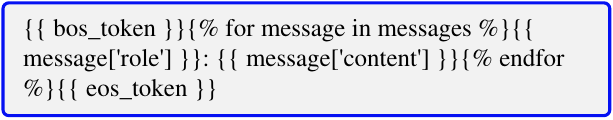}
    \label{fig:chat_template}
}

\subfloat[Message template.]{
    \includegraphics[width=0.75\columnwidth]{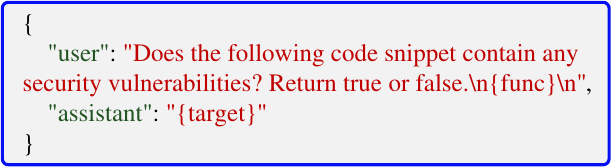}
    \label{fig:message_template}
}

\subfloat[Message template for GRACE.]{
    \includegraphics[width=0.75\columnwidth]{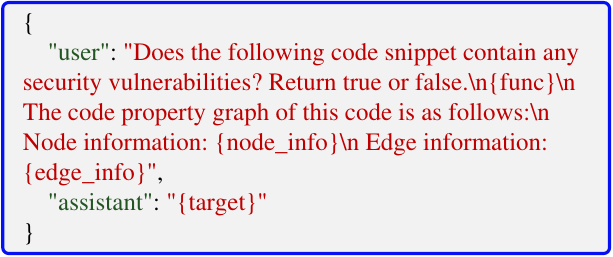}
    \label{fig:message_template_grace}
}

\caption{Templates used in the experiments.}
\label{fig:templates}
\end{figure}

To ensure consistency across experiments, the chat template is standardized, as shown in Figure \ref{fig:chat_template}. Additionally, as described in the previous sections, the following seven methods are selected as baselines for comparison with CGP-Tuning:
\begin{enumerate}
    \item \textbf{REVEAL \cite{chakraborty2022reveal}}: A graph neural network-based method begins by pre-training a graph neural network, then freezes it and adds a feature representation model at the end. The method then fine-tunes this feature representation model via contrastive learning.
    
    \item \textbf{Zero-Shot Prompting \cite{huang2024promptengineering}}: Utilizes a manually crafted prompt to instruct a pre-trained code LLM to classify whether given source code is vulnerable in a zero-shot manner. It leverages the in-context learning capabilities of the code LLM. The prompt, as shown in Figure \ref{fig:message_template}, provides the code LLM with source code accompanied by clear, task-oriented instructions.
    
    \item \textbf{GRACE \cite{lu2024grace}}: Building upon the concept of prompt engineering \cite{huang2024promptengineering}, this method describes the node and edge information of the code property graph using text. This information serves as additional context and is provided along with the task instruction and source code to the pre-trained code LLM, as shown in Figure \ref{fig:message_template_grace}.
    
    \item \textbf{Prompt Tuning \cite{lester2021prompttuning}}: A state-of-the-art soft prompt tuning method that uses a discrete textual prompt to initialize a set of trainable soft prompt embeddings. These embeddings are prepended to the word embeddings of the corresponding input as an enhancement to guide the code LLM toward the desired behavior. In the following experiments, 32 soft prompt embeddings are used, with a persona-based prompt pattern \cite{white2023promptpatterncatalogenhance} as initialization: “You are a highly skilled code auditor specializing in identifying vulnerabilities.".
    
    \item \textbf{G-Retriever \cite{he2024gretriever}}: A projector-based, graph-enhanced soft prompt tuning method. A graph neural network is first used to encode the code property graph into final node embeddings, which are subsequently aggregated into a graph embedding. A projector then maps the graph embedding into the latent space of the code LLM.
    
    \item \textbf{GNP \cite{tian2024gnp}}: A cross-attention-based, graph-enhanced soft prompt tuning method. A graph neural network first encodes the code property graph into final node embeddings, followed by self-attention to capture node significance and cross-attention to model graph-text interactions. The resulting aligned node embeddings are aggregated into an aligned graph embedding and projected into the code LLM's latent space via a projector.
    
    \item \textbf{Low-rank adaptation (LoRA) \cite{hu2022lora}}: Unlike soft prompt tuning, which keeps the pre-trained code LLM entirely frozen, LoRA enables lightweight adaptation by using low-rank matrices to approximate updates to selected weight matrices. It can also be used in conjunction with soft prompt tuning.
\end{enumerate}

The retrieval-augmented component in GRACE, G-Retriever, and GNP is removed to ensure a fair comparison. This component retrieves semantically similar samples as additional context, which is not the focus of this paper.

For prompt tuning, G-Retriever, GNP, the proposed CGP-Tuning, and LoRA, AdamW is used as the optimizer. The initial learning rate is set to 5e-5 with a linear learning rate scheduler, a batch size of one, and an accumulated batch size of 16. All methods are fine-tuned for a single epoch. The message template used to format the data is the same as the one used in zero-shot prompting, as shown in Figure \ref{fig:message_template}. 

For REVEAL, the optimizer and batch size are set to AdamW and 16, respectively, to maintain consistency with other baselines. The remaining hyperparameters, including the number of epochs, learning rate, and early stopping configurations used during pre-training and contrastive learning, are consistent with those reported in the original paper \cite{chakraborty2022reveal}.

For graph-enhanced soft prompt tuning methods, including G-Retriever, GNP, and the proposed CGP-Tuning, the same structure of the graph encoder is used as introduced in Section \ref{sec:graph_encoder}, employing four basic blocks with a dropout rate of 0.1. The number of attention heads in the GAT layer is set to 8, and the maximum number of nodes in the pooling layer is set to 4,096. REVEAL also adopts the same structure for a fair comparison. The structure of the projector for G-Retriever and GNP follows their respective original papers \cite{he2024gretriever, tian2024gnp}.

In CGP-Tuning, the dimension of the edge type embedding (i.e., $d_{t_e}$) is set to 128, and 32 soft prompt embeddings are employed to maintain consistency with prompt tuning. The initialization of the soft prompt embeddings and the shared weight matrices in multi-head attention follows the default settings of \texttt{torch.nn.Embedding} and \texttt{torch.nn.Linear}, respectively. For both GNP and CGP-Tuning, the number of attention heads is set to 32, with a dropout rate of 0.1.

For LoRA, the rank is set to 4 and $\alpha$ is set to 8, with the fine-tuned layers being the weight matrices of the query and value projection layers. When combined with CGP-Tuning, the weights from CGP-Tuning are loaded and frozen, and only the LoRA components are fine-tuned.

\subsection{Implementation Details}
This section outlines the techniques and configurations employed during the implementation. Figure \ref{fig:templates} depicts the templates mentioned in the previous sections. To prevent out-of-memory errors, the maximum number of tokens is set to 4,096 during training and 16,000 during testing. Gradient checkpointing and BFloat16 precision are employed to minimize memory consumption in both phases, enabling code LLMs to handle these large amounts of tokens.

All experiments were conducted using a single NVIDIA H100 GPU, and the key packages are as follows: Python v3.11.9, Transformers v4.46.1\footnote{\url{https://github.com/huggingface/transformers}}, PEFT v0.13.2\footnote{\url{https://github.com/huggingface/peft}}, TRL v0.11.0\footnote{\url{https://github.com/huggingface/trl}}, Torch v2.4.0\footnote{\url{https://pytorch.org/}}, Pytorch-Lightning v2.4.0\footnote{\url{https://github.com/Lightning-AI/pytorch-lightning}}, CUDA Toolkit v11.8.0, and cuDNN v8.9.2.26.

\section{Results and Analysis}
This section provides an analysis and summary of the results from CGP-Tuning and other methods, with all values rounded to four decimal places for clarity.

\subsection{RQ1: Does CGP-Tuning outperform both traditional methods and state-of-the-art graph-enhanced soft prompt tuning methods in vulnerability detection?}
\begin{table*}[ht]
\centering
\caption{Performance Evaluation Results of Different Methods}
\label{tab:performance}
\resizebox{\textwidth}{!}{ 
\begin{tabular}{@{}l|cccc|cccc|cccc@{}}
\toprule
\multicolumn{1}{c}{\textbf{}} & \multicolumn{4}{c|}{\textbf{CodeLlama}} & \multicolumn{4}{c|}{\textbf{CodeGemma}} & \multicolumn{4}{c}{\textbf{Qwen2.5-Coder}} \\
\cmidrule(lr){2-5} \cmidrule(lr){6-9} \cmidrule(lr){10-13}
\multicolumn{1}{c}{\textbf{Methods}} & \textbf{Accuracy} & \textbf{Precision} & \textbf{Recall} & \textbf{F1-score} & \textbf{Accuracy} & \textbf{Precision} & \textbf{Recall} & \textbf{F1-score} & \textbf{Accuracy} & \textbf{Precision} & \textbf{Recall} & \textbf{F1-score} \\
\midrule
REVEAL & 0.5000 & 0.2500 & 0.5000 & 0.3333 & 0.5000 & 0.2500 & 0.5000 & 0.3333 & 0.5000 & 0.2500 & 0.5000 & 0.3333 \\
Zero-Shot Prompting & 0.5060 & 0.6073 & 0.5060 & 0.3535 & 0.4681 & 0.4605 & 0.4681 & 0.4410 & 0.5275 & 0.5459 & 0.5275 & 0.4746 \\
GRACE & 0.5958 & 0.6427 & 0.5958 & 0.5596 & 0.4020 & 0.4003 & 0.4020 & 0.3995 & 0.5692 & 0.6333 & 0.5692 & 0.5103 \\
Prompt Tuning & 0.5901 & 0.6539 & 0.5901 & 0.5426 & 0.5113 & 0.5121 & 0.5113 & 0.5026 & 0.5758 & 0.6431 & 0.5758 & 0.5192 \\
G-Retriever & 0.6266 & 0.6526 & 0.6266 & 0.6099 & 0.5692 & 0.6077 & 0.5692 & 0.5269 & 0.5991 & 0.6017 & 0.5991 & 0.5965 \\
GNP & 0.6356 & 0.6359 & 0.6356 & 0.6355 & 0.5755 & 0.5769 & 0.5755 & 0.5736 & 0.6076 & 0.6241 & 0.6076 & 0.5942 \\
CGP-Tuning & 0.6582 & 0.6590 & 0.6582 & 0.6577 & \textbf{0.6219} & 0.6362 & \textbf{0.6219} & \textbf{0.6118} & 0.6507 & 0.6697 & 0.6507 & 0.6407 \\
LoRA & 0.6222 & \textbf{0.6920} & 0.6222 & 0.5844 & 0.5870 & 0.6493 & 0.5870 & 0.5390 & 0.6757 & 0.6761 & 0.6757 & 0.6756 \\
LoRA + CGP-Tuning & \textbf{0.6678} & 0.6678 & \textbf{0.6678} & \textbf{0.6678} & 0.6021 & \textbf{0.6610} & 0.6021 & 0.5621 & \textbf{0.6884} & \textbf{0.6899} & \textbf{0.6884} & \textbf{0.6877} \\
\bottomrule
\end{tabular}
}
\vspace{1mm} 
\begin{minipage}{0.9\textwidth}
\vspace{1mm}
\footnotesize \textit{Note: The best results are highlighted in \textbf{bold}. The rest of the tables in this paper follow the same convention.}
\end{minipage}
\end{table*}

To evaluate whether the proposed CGP-Tuning achieves the best overall performance in vulnerability detection, Table \ref{tab:performance} reports the evaluation results across various methods. The comparison between REVEAL and zero-shot prompting shows that REVEAL only matches the performance of weaker code LLMs, such as CodeGemma, in terms of accuracy and recall. However, it lags significantly behind more advanced code LLMs like CodeLlama and Qwen2.5-Coder, highlighting the superiority of code LLMs over traditional methods in vulnerability detection.

Moving to methods powered by code LLMs, the results from zero-shot prompting suggest that, even without fine-tuning, these code LLMs can still identify some vulnerabilities due to their in-context learning abilities. However, the effectiveness of this capability varies significantly between different code LLMs. In the context of vulnerability detection, CodeLlama and Qwen2.5-Coder significantly outperform CodeGemma, achieving accuracies of approximately 51\% and 53\%, respectively, compared to CodeGemma's 47\%, which is worse than random guessing.

The variation in in-context learning capabilities among code LLMs further impacts how well GRACE performs. When used with CodeLlama and Qwen2.5-Coder, GRACE surpasses zero-shot prompting because these models have strong in-context learning abilities that help them extract meaningful insights from the graph-based structural information within the context, thereby enhancing their vulnerability detection performance. In contrast, CodeGemma has weaker in-context learning capabilities, so adding structural information actually confuses the model and leads to significantly worse performance than simple zero-shot prompting.

The performance of the traditional soft prompt tuning method, prompt tuning, is also heavily affected by the in-context learning capabilities of the underlying code LLMs. For code LLMs that have strong in-context learning abilities, prompt tuning provides only marginal benefits. With CodeLlama, prompt tuning shows better precision than GRACE but performs worse on other metrics. For Qwen2.5-Coder, prompt tuning slightly outperforms GRACE across all metrics by roughly half a percentage point. These modest improvements occur because both CodeLlama and Qwen2.5-Coder can already effectively interpret and use the structural information that GRACE provides in context, making the performance gap between the two approaches relatively small. 

The picture changes dramatically with CodeGemma, a model with weaker in-context learning capabilities. Here, prompt tuning delivers substantial improvements over GRACE across all metrics. The learned soft prompt embeddings help compensate for CodeGemma's limited ability to understand context effectively. However, even with these improvements, traditional prompt tuning still underperforms compared to graph-enhanced soft prompt tuning methods. This limitation stems from prompt tuning's inability to incorporate the graph-based structural information inherent in source code.

Among graph-enhanced soft prompt tuning methods, G-Retriever uses a projector-based cross-modal alignment module with a linear computational cost proportional to $|V|$. It shows strong precision on code LLMs like CodeLlama and CodeGemma, but its limited modeling of graph-text interactions reduces performance on other metrics compared to GNP. However, the cross-attention-based module used in GNP comes at a significantly higher computational cost scaling with $|V|^2 + |V| \times N + |V|$. The proposed CGP-Tuning balances both aspects, achieving a linear computational cost of $|V| + N$ while effectively incorporating graph-text interactions and semantic information in code property graphs. This enables CGP-Tuning to achieve the best overall performance.

Specifically, compared to zero-shot prompting, CGP-Tuning achieves substantial performance improvements across all code LLMs. CodeGemma, which has the weakest in-context learning ability, benefits from an average improvement of 16.5 percentage points across all metrics. CodeLlama and Qwen2.5-Coder, despite their stronger in-context learning capabilities, still gain 16.4 and 13.4 percentage points respectively. These consistent improvements across all evaluated code LLMs show no significant diminishing returns, even for models with strong base performance due to their superior in-context learning abilities. This shows that the benefits of CGP-Tuning are model-agnostic.

Compared to traditional soft prompt tuning method and latest graph-enhanced variants, CGP-Tuning continuously delivers substantial improvements: nearly eight percentage points on average over the worst-performing prompt tuning, five percentage points over G-Retriever, and four percentage points compared to the best-performing GNP. Even compared to LoRA, an advanced parameter-efficient fine-tuning technique that updates the internal weight matrices of code LLMs, the proposed CGP-Tuning consistently outperforms it across most evaluated code LLMs (i.e., CodeLlama and CodeGemma), while keeping the pre-trained code LLMs completely frozen.

Furthermore, integrating LoRA with CGP-Tuning yields additional performance gains, with particularly notable improvements observed in code LLMs possessing strong in-context learning capabilities, such as CodeLlama and Qwen2.5-Coder. Notably, Qwen2.5-Coder achieves optimal performance in this experiment when augmented with both LoRA and CGP-Tuning, reaching about 69\% across all metrics. Even for code LLMs with weaker in-context learning abilities like CodeGemma, the combination of LoRA and CGP-Tuning delivers measurable improvements over using LoRA alone. These findings conclusively demonstrate the versatility and effectiveness of CGP-Tuning in enhancing the performance of code LLMs for vulnerability detection.

\begin{rqbox}[label={rq:rq1}]{Answer to RQ1}
CGP-Tuning's improvement is model-agnostic. It consistently outperforms traditional methods, soft prompt tuning methods, and the latest graph-enhanced variants. Even compared to LoRA, a technique that updates the internal weight matrices of code LLMs, CGP-Tuning still achieves optimal performance on CodeLlama and CodeGemma, and can further enhance code LLMs' performance when combined with LoRA. These results demonstrate that CGP-Tuning is a versatile and effective solution for vulnerability detection.
\end{rqbox}

\subsection{RQ2: Does incorporating the cross-modal alignment module, type-aware embeddings, and positional embeddings effective in improving the performance of CGP-Tuning?}
\begin{table*}[ht]
\centering
\caption{Ablation Experiment Results on CGP-Tuning}
\label{tab:ablation_study}
\resizebox{\textwidth}{!}{ 
\begin{tabular}{@{}l|cccc|cccc|cccc@{}}
\toprule
\multicolumn{1}{c}{\textbf{}} & \multicolumn{4}{c|}{\textbf{CodeLlama}} & \multicolumn{4}{c|}{\textbf{CodeGemma}} & \multicolumn{4}{c}{\textbf{Qwen2.5-Coder}} \\ 
\cmidrule(lr){2-5} \cmidrule(lr){6-9} \cmidrule(lr){10-13}
\multicolumn{1}{c}{\textbf{Methods}} & \textbf{Accuracy} & \textbf{Precision} & \textbf{Recall} & \textbf{F1-score} & \textbf{Accuracy} & \textbf{Precision} & \textbf{Recall} & \textbf{F1-score} & \textbf{Accuracy} & \textbf{Precision} & \textbf{Recall} & \textbf{F1-score} \\ 
\midrule
CGP-Tuning w/o multi-head attention & 0.5426 & 0.5429 & 0.5426 & 0.5416 & 0.5089 & 0.5095 & 0.5089 & 0.5010 & 0.5392 & 0.5702 & 0.5392 & 0.4819 \\
CGP-Tuning w/o projector & 0.4334 & 0.4076 & 0.4334 & 0.3909 & 0.4768 & 0.3574 & 0.4768 & 0.3383 & 0.5397 & 0.5751 & 0.5397 & 0.4782 \\
CGP-Tuning w/o cross-modal alignment & 0.4752 & 0.4050 & 0.4752 & 0.3563 & 0.4700 & 0.4659 & 0.4700 & 0.4537 & 0.5123 & 0.5268 & 0.5123 & 0.4358 \\
CGP-Tuning w/o node type embeddings & 0.6527 & 0.6550 & 0.6527 & 0.6515 & 0.6055 & 0.6223 & 0.6055 & 0.5914 & 0.6543 & 0.6648 & 0.6543 & 0.6487 \\
CGP-Tuning w/o edge type embeddings & 0.6606 & 0.6608 & 0.6606 & 0.6604 & 0.6141 & 0.6305 & 0.6141 & 0.6016 & 0.6551 & 0.6693 & 0.6551 & 0.6477 \\
CGP-Tuning w/o positional embeddings & 0.6047 & 0.6363 & 0.6047 & 0.5804 & 0.6052 & 0.6089 & 0.6052 & 0.6018 & 0.6146 & 0.6708 & 0.6146 & 0.5801 \\
CGP-Tuning & \textbf{0.6627} & \textbf{0.6629} & \textbf{0.6627} & \textbf{0.6626} & \textbf{0.6154} & \textbf{0.6331} & \textbf{0.6154} & \textbf{0.6022} & \textbf{0.6572} & \textbf{0.6713} & \textbf{0.6572} & \textbf{0.6500} \\
\bottomrule
\end{tabular}
}
\end{table*}

To assess the effectiveness of the cross-modal alignment module, type-aware embeddings, and positional embeddings in the proposed CGP-Tuning, this section presents detailed ablation experiments by ablating one component at a time. The results are derived from the validation dataset to avoid any leakage into the test dataset in other experiments.

As presented in Table \ref{tab:ablation_study}, ablating each component of CGP-Tuning consistently degrades performance across all evaluated code LLMs, demonstrating the effectiveness of each component. The most significant performance drop occurs when the entire cross-modal alignment module is ablated. This configuration skips step \circled{10} and uses the soft prompt embeddings at step \circled{2} directly to replace step \circled{11} in Figure \ref{fig:cgp_tuning}, leading to accuracy drops of approximately 19, 15, and 14 percentage points on CodeLlama, CodeGemma, and Qwen2.5-Coder, respectively. This decline occurs because plain soft prompt embeddings cannot bridge the semantic gap between graph and text features without the multi-head attention mechanism, nor can they properly integrate with the code LLM's latent space without the projection.

To evaluate the contributions of the multi-head attention mechanism and the projector in the cross-modal alignment module (Figure \ref{fig:cgp_tuning}, step \circled{10}), additional ablation experiments were conducted. Specifically, the projector was bypassed by directly using the output of the multi-head attention layers (i.e., graph-aligned features $H_5$) to replace step \circled{11} in Figure \ref{fig:cgp_tuning}. The multi-head attention mechanism was bypassed by using the soft prompt embeddings at step \circled{2} as the input to equation \ref{eq:ffn} and the corresponding output to replace step \circled{11} in Figure \ref{fig:cgp_tuning}. As shown in Table \ref{tab:ablation_study}, ablating the projector causes a more pronounced overall performance decline compared to ablating the multi-head attention mechanism. This indicates that feature alignment through multi-head attention alone is insufficient for obtaining usable aligned features; the aligned features must be projected into a representation that code LLMs can interpret.

In type-aware embeddings, ablating node type embeddings involves using only the positional embeddings from step \circled{7} as node embeddings $H_1$, along with the edge type embeddings in step \circled{6}, both serving as input to step \circled{8} of Figure \ref{fig:cgp_tuning}. This results in a notable performance decline, with an average drop of approximately one percentage point across all metrics.  Ablating edge type embeddings is done by using only the sum of the node type embeddings and the positional embeddings in steps \circled{5} and \circled{7} as node embeddings $H_1$, which serves as the input to step \circled{8} of Figure \ref{fig:cgp_tuning}. This ablation has a comparatively smaller impact on performance. The greater performance drop from removing node type embeddings likely arises from the fact that edge types can often be inferred from the types of the connected nodes, whereas the reverse is less feasible.

Compared to type-aware embeddings, positional embeddings play a more critical role in graph representation, as evidenced by a larger performance drop when they are ablated. This ablation is implemented by using only node type embeddings from step \circled{5} as $H_1$, along with edge type embeddings from step \circled{6}, as input to step \circled{8} in Figure \ref{fig:cgp_tuning}. This finding highlights the importance of positional embeddings in ensuring that each node receives a unique representation, which type-aware embeddings alone cannot guarantee. 

These results demonstrate that both node and edge types in the code property graph carry rich semantic information, and that type-aware embeddings and positional embeddings are effective in capturing these semantics to help code LLMs improve their understanding of code structures.

\begin{rqbox}[label={rq:rq2}]{Answer to RQ2}
Incorporating the cross-modal alignment module, type-aware embeddings, and positional embeddings is effective in improving the performance of CGP-Tuning. Ablation results show that ablating any component consistently degrades performance across all evaluated code LLMs. The cross-modal alignment module has the most significant impact, particularly the projector. Type-aware embeddings also contribute meaningfully, with node type embeddings more influential than edge type embeddings. Positional embeddings are crucial for distinguishing node representations, causing the largest drop among embedding components when ablated.
\end{rqbox}

\subsection{RQ3: Does CGP-Tuning reduce computational costs in its cross-modal alignment module at the expense of performance when processing long source code?}
\begin{table*}[ht]
\centering
\caption{Performance Evaluation Results of Different Methods Under Long Source Code}
\label{tab:long_source_code}
\resizebox{\textwidth}{!}{ 
\begin{tabular}{@{}l|cccc|cccc|cccc@{}}
\toprule
\multicolumn{1}{c}{\textbf{}} & \multicolumn{4}{c|}{\textbf{CodeLlama}} & \multicolumn{4}{c|}{\textbf{CodeGemma}} & \multicolumn{4}{c}{\textbf{Qwen2.5-Coder}} \\ 
\cmidrule(lr){2-5} \cmidrule(lr){6-9} \cmidrule(lr){10-13}
\multicolumn{1}{c}{\textbf{Methods}} & \textbf{Accuracy} & \textbf{Precision} & \textbf{Recall} & \textbf{F1-score} & \textbf{Accuracy} & \textbf{Precision} & \textbf{Recall} & \textbf{F1-score} & \textbf{Accuracy} & \textbf{Precision} & \textbf{Recall} & \textbf{F1-score} \\ 
\midrule
Zero-Shot Prompting & 0.1811 & 0.7395 & 0.1811 & 0.1608 & 0.2598 & 0.7709 & 0.2598 & 0.2841 & 0.2362 & 0.6947 & 0.2362 & 0.2731 \\
GRACE & 0.5906 & \textbf{0.8194} & 0.5906 & 0.6612 & 0.3268 & 0.7620 & 0.3268 & 0.3863 & 0.4764 & \textbf{0.7797} & 0.4764 & 0.5588 \\
Prompt Tuning & \textbf{0.8780} & 0.7708 & \textbf{0.8780} & \textbf{0.8209} & 0.7126 & 0.7688 & 0.7126 & 0.7383 & 0.8583 & 0.7621 & 0.8583 & 0.8074 \\
G-Retriever & \textbf{0.8780} & 0.7708 & \textbf{0.8780} & \textbf{0.8209} & 0.5984 & 0.7585 & 0.5984 & 0.6608 & 0.6535 & 0.7777 & 0.6535 & 0.7041 \\
GNP & \textbf{0.8780} & 0.7708 & \textbf{0.8780} & \textbf{0.8209} & 0.6339 & 0.7616 & 0.6339 & 0.6863 & 0.8504 & 0.7796 & 0.8504 & 0.8093 \\
CGP-Tuning & \textbf{0.8780} & 0.7708 & \textbf{0.8780} & \textbf{0.8209} & \textbf{0.7362} & 0.7520 & \textbf{0.7362} & \textbf{0.7440} & \textbf{0.8740} & 0.7639 & \textbf{0.8740} & \textbf{0.8153} \\
LoRA & 0.8504 & 0.7678 & 0.8504 & 0.8070 & 0.6496 & \textbf{0.7796} & 0.6496 & 0.7008 & \textbf{0.8740} & 0.7639 & \textbf{0.8740} & \textbf{0.8153} \\
LoRA + CGP-Tuning & \textbf{0.8780} & 0.7708 & \textbf{0.8780} & \textbf{0.8209} & 0.6811 & 0.7767 & 0.6811 & 0.7214 & \textbf{0.8740} & 0.7639 & \textbf{0.8740} & \textbf{0.8153} \\
\bottomrule
\end{tabular}
}
\end{table*}

To evaluate whether CGP-Tuning reduces computational costs in its cross-modal alignment module at the expense of performance when processing long source code, this section examines performance differences among various methods when the input source code is very long. Specifically, samples are sorted by their number of tokens, and the top 7\% of the longest samples are retained. Due to differences in tokenizers and vocabularies across different code LLMs, the average number of tokens is approximately 6,300 for CodeLlama, 5,500 for CodeGemma, and 4,600 for Qwen2.5-Coder. Since REVEAL is based on graph neural networks and does not use tokenizers and vocabularies, making it incompatible with this experiment, it is excluded from the analysis.

As presented in Table \ref{tab:long_source_code}, the results of zero-shot prompting reveal that all evaluated code LLMs exhibit reduced effectiveness in identifying vulnerabilities within long source code. Compared to Table \ref{tab:performance}, performance is nearly halved across most metrics. However, their precision is significantly higher. This pattern suggests that code LLMs adopt a more conservative approach when handling long source code: they correctly identify fewer vulnerabilities but demonstrate greater certainty in their identifications.

The results of GRACE demonstrate that incorporating structural information from code property graphs as context significantly enhances code LLM performance compared to zero-shot prompting. For code LLMs with strong in-context learning capabilities, such as CodeLlama and Qwen2.5-Coder, GRACE increases accuracy from 18\% to 59\% and 24\% to 48\%, respectively, nearly doubling or even tripling performance. Even for CodeGemma, which exhibits weaker in-context learning capabilities, accuracy still improves from 26\% to 33\%. These findings indicate that when detecting vulnerabilities in long source code, integrating graph-based structural information as context allows code LLMs to identify more vulnerabilities (improving recall) while maintaining or improving precision.

Compared to zero-shot prompting and GRACE, traditional soft prompt tuning and the latest graph-enhanced variants improve recall at a moderate cost to precision. These results demonstrate the effectiveness of learned soft prompt embeddings in enhancing code LLMs to accurately detect more vulnerabilities in long source code. Notably, the proposed CGP-Tuning boosts performance on all metrics by an average of 34.9 percentage points for CodeGemma, 52.1 percentage points for CodeLlama, and 47.2 percentage points for Qwen2.5-Coder. Unlike the regular scenario in Table \ref{tab:performance}, in the long source code scenario, the larger gains for CodeLlama and Qwen2.5-Coder suggest that models with strong in-context learning capabilities (i.e., high base performance) benefit more from CGP-Tuning than models like CodeGemma.

Traditional soft prompt tuning and latest graph-enhanced variants perform identically on CodeLlama. However, with models like CodeGemma and Qwen2.5-Coder, more noticeable differences emerge. In some cases, prompt tuning even outperforms certain graph-enhanced soft prompt tuning methods, like G-Retriever and GNP, though CGP-Tuning remains superior in general. The performance gap likely stems from G-Retriever and GNP pooling graph features into a single graph embedding before projection, leading to information loss that becomes especially problematic with long source code.

CGP-Tuning takes a different approach by using soft prompt embeddings as global context that interacts with both graph and text features. By generating 32 soft prompt embeddings, similar to prompt tuning, and utilizing type-aware embeddings to capture semantic information within the code property graph, it mitigates information loss. This design enables CGP-Tuning to consistently achieve the best overall performance across all evaluated code LLMs.

LoRA, despite its advantage of updating the internal weight matrices of code LLMs, still demonstrates inferior overall performance on long source code compared to soft prompt tuning methods. This suggests that learned soft prompt embeddings are better at helping code LLMs detect vulnerabilities in long source code. Integrating LoRA with CGP-Tuning can address this limitation for CodeLlama and Qwen2.5-Coder, though for CodeGemma, this combination only beats standalone LoRA and still falls short of CGP-Tuning alone.

Interestingly, several tuning-based methods achieve identical performance when applied to CodeLlama and Qwen2.5-Coder, but show more variation with CodeGemma. This difference likely stems from the inherent performance bottlenecks of the underlying code LLMs. When handling long source code, the improvements these methods can bring are largely constrained by the capabilities of the underlying code LLMs. CodeLlama and Qwen2.5-Coder reach this bottleneck faster because their strong in-context learning abilities provide better base performance, while CodeGemma's weaker in-context learning abilities leave more room for performance variation.

\begin{rqbox}[label={rq:rq3}]{Answer to RQ3}
CGP-Tuning does not reduce computational costs at the expense of performance when handling long source code. Rather, it effectively preserves more semantic information from the code property graph without excessive information loss in its graph-enhanced soft prompt embeddings, consistently delivering superior overall performance.
\end{rqbox}

\subsection{RQ4: Does CGP-Tuning's computational efficiency consistent with the theoretical expectations?}

To verify whether CGP-Tuning demonstrates computational efficiency as theoretically expected, Table \ref{tab:total_inference_time} presents the total inference time (in minutes) for various methods across all evaluated code LLMs on the test subset.

It is observed that, owing to REVEAL's small parameter size in comparison to modern code LLMs, it achieves the shortest total inference time, completing execution in under 30 seconds. However, as shown in Table \ref{tab:performance}, this efficiency comes at the cost of performance, with REVEAL exhibiting the lowest performance. The results of zero-shot prompting reveal that subtle variations in model architecture and tokenizer design contribute to the observed discrepancies in total inference times across different code LLMs. Nonetheless, the overall inference latency remains reasonable, with all code LLMs completing within four minutes. GRACE incurs the highest total inference time due to its use of a contextualized code property graph combined with the quadratic complexity of self-attention. Notably, GRACE with CodeGemma demonstrates the longest inference duration, exceeding half an hour.

The traditional soft prompt tuning method, prompt tuning, increases total inference time by about one percent on average compared to zero-shot prompting, as it only adds a fixed number of soft prompt embeddings during inference. Among graph-enhanced soft prompt tuning methods, CGP-Tuning exhibits a total inference time that falls between those of G-Retriever and GNP, which is consistent with its theoretical efficiency ranking. Compared to zero-shot prompting, which is the fastest LLM-driven method as it introduces no extra components, CGP-Tuning adds only 39 seconds on average to the total inference time. This additional complexity is negligible considering the significant performance improvement shown by CGP-Tuning in Table \ref{tab:performance}.

LoRA's total inference time remains nearly identical to that of zero-shot prompting, as its low-rank matrices are lightweight and introduce minimal overhead during inference. When LoRA is combined with CGP-Tuning, the overall increase in total inference time is consistent with using CGP-Tuning alone, which aligns with LoRA's theoretical efficiency. 

\begin{rqbox}[label={rq:rq4}]{Answer to RQ4}
The empirical results confirm that CGP-Tuning achieves computational efficiency that matches theoretical expectations. Its total inference time falls between that of G-Retriever and GNP, aligning closely with the expected efficiency. Moreover, integrating CGP-Tuning with LoRA incurs minimal additional overhead, requiring less than one minute on average. Overall, CGP-Tuning can substantially improve the vulnerability detection capabilities of code LLMs without sacrificing practical inference speed.
\end{rqbox}

\begin{table}[ht]
\centering
\caption{Total Inference Time (in Minutes) of Different Methods}
\label{tab:total_inference_time}
\begin{tabular}{@{}l|ccc@{}}
\toprule
\multicolumn{1}{c}{\textbf{Methods}} & \textbf{CodeLlama} & \textbf{CodeGemma} & \textbf{Qwen2.5-Coder} \\
\midrule
REVEAL & \textbf{0.4012} & \textbf{0.4012} & \textbf{0.4012} \\
Zero-Shot Prompting & 3.6184 & 3.7336 & 2.5301 \\
GRACE & 29.7277 & 31.7243 & 24.3495 \\
Prompt Tuning & 3.6868 & 3.7841 & 2.5582 \\
G-Retriever & 4.1082 & 4.2009 & 3.0527 \\
GNP & 4.3805 & 4.4596 & 3.3670 \\
CGP-Tuning & 4.2753 & 4.3435 & 3.2118 \\
LoRA & 3.6502 & 3.7380 & 2.7395 \\
LoRA + CGP-Tuning & 4.2663 & 4.3594 & 3.4739 \\
\bottomrule
\end{tabular}
\end{table}

\section{Related Work}
This section summarizes the literature related to the topic of this work, focusing on how traditional methods leverage graph-based structural information and how the latest LLMs incorporate such information.

\subsection{Graph-Based Structural Info. in Traditional Methods}
Vulnerability detection aims to identify security flaws within source code. A common approach involves constructing code graphs that capture the structural and semantic properties of the code. Static analysis methods typically rely on these graphs to model control and data flow information \cite{sui2016svf}, which are useful for detecting specific issues like memory leaks \cite{sui2012memoryleak} and null pointer dereference \cite{tomassi2021nullpointer}. However, due to their rule-based nature, static analysis methods often struggle to detect novel or complex vulnerabilities that deviate from predefined patterns.

To enhance generalization, recent advancements have focused on deep learning-based approaches \cite{chakraborty2022reveal}, particularly by using pre-trained code language models such as CodeBERT \cite{feng2020codebert} and CodeT5 \cite{wang2021codet5}, which can be fine-tuned to classify source code as either vulnerable or secure \cite{thapa2022vulcodebert}. However, these models often treat code as plain text, failing to capture the rich, graph-based structural information inherent in source code.

To incorporate such structural information, one approach is to traverse the code graph, flatten it into a sequence, and pre-train the model with a link prediction task to capture adjacency information, as seen in GraphCodeBERT \cite{guo2021graphcodebert}. GraphCodeBERT incorporates data flow graphs during pre-training and can then be fine-tuned for vulnerability detection \cite{wang2024scl}. However, recent insights from scaling laws \cite{kaplan2020scalinglawsneurallanguage} suggest that traditional code language models are constrained by both the model size and the scale of the dataset used during pre-training, which limits their overall capabilities.

Another approach involves the use of graph neural networks \cite{kipf2017gcn, velickovic2018gat}. Unlike GraphCodeBERT, graph neural networks can directly incorporate adjacency information from code graphs. By constructing various graph representations (e.g., control flow graphs, data flow graphs, and abstract syntax trees) from source code, they can effectively capture the graph-based structural information necessary to identify potential vulnerabilities \cite{cheng2021deepwukong, xiao2024enhancedgraph, qiu2024multiplegraph, chakraborty2022reveal}. However, graph neural networks face challenges like over-smoothing, which limits their network depth and ability to capture complex dependencies \cite{wu2023oversmoothing}. Additionally, they are less effective at capturing long-range contextual information compared to code language models.

\subsection{Graph-Based Structural Info. in Large Language Models}
With the rise of LLMs, recent research has focused on integrating graph-based structural information with LLMs. One approach is to describe the graph's adjacency information in natural language, leveraging the in-context learning ability of LLMs to incorporate this graph-based structural information \cite{fatemi2024talk}. GRACE \cite{lu2024grace} adopts this strategy by first representing the source code with code property graphs \cite{fabian2014cpg}, then describing the adjacency information in natural language as additional context for LLMs, which has been shown to improve vulnerability detection performance. However, this approach heavily relies on the in-context learning capabilities of LLMs and is computationally expensive, primarily due to the quadratic complexity of the self-attention mechanism \cite{vaswani2017attention}.

Another way is through graph neural networks \cite{tang2024graphgpt, ma2024xrec, he2024gretriever, tian2024gnp} or flattened graph sequences \cite{chen2024llaga} to extract graph features, and leverage graph-enhanced soft prompt tuning to interpret the extracted graph features into soft prompt embeddings. Since the LLM is frozen in this process, cross-modal alignment modules are essential to incorporate the graph features into the soft prompt embeddings. The design of these cross-modal alignment modules can be broadly categorized into projector-based methods and cross-attention-based methods. 

In projector-based methods, the feed-forward networks are typically used for projecting the extracted graph features. For node-level tasks, the projector maps node embeddings individually \cite{chen2024llaga, tang2024graphgpt, ma2024xrec}, while for graph-level tasks, the node embeddings are pooled into a single graph embedding, which is then projected into the LLM's latent space \cite{he2024gretriever}. Projector-based methods are highly efficient but neglect the correlations between graph features and text features.

On the other hand, cross-attention-based methods \cite{tian2024gnp} incorporate the cross-attention mechanisms to model graph-text interactions more effectively, but they often incur high computational costs due to the need to compute and store large attention score matrices. Moreover, existing graph-enhanced soft prompt tuning methods are designed for general graph-related tasks \cite{chen2024llaga, tang2024graphgpt, ma2024xrec, he2024gretriever, tian2024gnp} and fail to capture the rich semantic information in code graphs. 

The proposed CGP-Tuning overcomes these limitations by modeling graph-text interactions while ensuring computational efficiency within its cross-modal alignment module. Additionally, it introduces type-aware embeddings to capture semantic information in code graphs, offering an improved solution for vulnerability detection with modern code LLMs.
\section{Limitations and Threats to Validity}
In this section, some major limitations and threats that could impact the validity of the results are discussed.

\begin{enumerate}
    \item \textbf{Exclusion of larger code LLMs}. The exclusion of larger code LLMs due to hardware constraints limits direct comparability with the highest-performing code LLMs available. This may affect the absolute performance ceilings, as scaling laws suggest larger models tend to achieve qualitatively better capabilities \cite{kaplan2020scalinglawsneurallanguage}. The tested Qwen2.5-Coder mitigates this problem, as its 7B version is more up-to-date than some older (but larger) code LLMs like the WizardCoder 15B version. Moreover, a recent study \cite{hui2024qwen25coder} shows its 7B version performs consistently better than many of the latest, larger code LLMs like DeepSeek-Coder-V2 16B version. Hence, the representativeness of the evaluation is ensured. Furthermore, the goal is to demonstrate the relative performance differences among different methods, for which the selected code LLMs are well-suited.
    
    \item \textbf{Limited scope of tested code LLMs}. As newer and more powerful code LLMs continue to emerge, evaluating all of them is infeasible due to the computational resources and manual effort required. Further, the absence of benchmark results for vulnerability detection on the latest code LLMs makes it challenging to identify those that are most suitable for this work. Although an empirical study of the latest code LLMs for vulnerability detection would be valuable, such a study lies beyond the scope of this paper. To ensure representativeness of the evaluation, this paper selects CodeLlama \cite{roziere2024codellama}, CodeGemma \cite{zhao2024codegemma}, and Qwen2.5-Coder \cite{hui2024qwen25coder}, which are among the latest, state-of-the-art code LLMs from industry and have been widely adopted in recent research \cite{lam2024llmrpg, yu2024codereview, nashaat2024codementor, sultana2024codevulnerabilitydetectioncomparative}. In addition, the implementation is open-sourced to facilitate the extension of CGP-Tuning to larger and newer code LLMs.
    
    \item \textbf{Representativeness of vulnerability detection dataset}. The validity of the results may be affected by the representativeness of the vulnerability detection dataset used in the evaluation. This paper does not utilize widely adopted vulnerability detection datasets from earlier research such as REVEAL \cite{chakraborty2022reveal}, as these datasets are somewhat outdated and more likely to contain samples that code LLMs may have already encountered during pre-training. Therefore, this paper selects the latest available DiverseVul dataset \cite{chen2023diversevul} at the time of implementation. However, the results may still be influenced by biases within the dataset.

    \item \textbf{Potential data contamination and duplication issues}. Recent studies \cite{dong2024generalization, lopez2025dataleakage4se} have noted that LLM evaluations can sometimes be affected by data contamination, where evaluation datasets may have been partially seen during pre-training. This includes data duplication where overlapping samples between pre-training and downstream tasks could moderately inflate evaluation metrics. As existing open-source code LLMs \cite{roziere2024codellama, zhao2024codegemma, hui2024qwen25coder} do not disclose their pre-training corpora, this issue is beyond our control. Moreover, recent contamination detection techniques focus on sequence generation \cite{dong2024generalization} and are not applicable to classification tasks.
    
    Nevertheless, even if absolute performance metrics are slightly inflated due to potential pre-training contamination, the relative improvements achieved by CGP-Tuning remain valid. The DiverseVul \cite{chen2023diversevul} dataset ensures that no data duplication occurs, which guarantees test set integrity as the test set had no overlap with the training set used for fine-tuning.  Additionally, the ablation study was conducted on the validation set without any test set leakage. Thus, since all methods are equally exposed to the same contamination risk, the comparative analysis remains fair.
    
    \item \textbf{Limitations of learning-based approaches}. This paper demonstrates the effectiveness of incorporating graph-based structural information to enhance code LLMs' vulnerability detection capabilities, but it does not focus on addressing common limitations inherent in learning-based approaches \cite{gao2024learning}, such as detecting out-of-distribution samples like previously unknown vulnerability types, or temporal bias where the model may underperform on vulnerabilities that emerge after the fine-tuning period. Future work could investigate techniques like continual learning or domain adaptation to improve the model's robustness to such challenges.

    \item \textbf{Dependence on labeled data}. Similar to other tuning-based approaches, CGP-Tuning relies on labeled data for supervised fine-tuning. This limitation can be addressed by utilizing publicly available datasets (e.g., DiverseVul) to fine-tune code LLMs, which can then be directly deployed in production. While labeled data tailored to developers' specific use cases would be preferable, it is not strictly necessary.
    
    \item \textbf{Limited maximum number of tokens}. Due to hardware limitations and the need for fairness in comparisons, this paper sets the maximum number of tokens to 16,000 during testing, based on hardware capacity. However, for certain exceptionally long samples (approximately 0.081\% of the samples in the original dataset, averaged across the three evaluated code LLMs), truncating the overflowed tokens may impact the evaluation of different methods on these samples. This limitation also restricts the applicability of code LLMs to the analysis of large-scale projects. It is important to note that this limitation is not unique to this paper but is commonly encountered by various methods in the field.
    
    \item \textbf{Limited scope of tested task}. The evaluation and analysis in this paper focus solely on vulnerability detection, which may restrict the generalizability of the findings to other code-related tasks. While the proposed CGP-Tuning proves effective for vulnerability detection, its performance in tasks like code generation or code search remains unexplored. These tasks often require distinct capabilities such as the ability to synthesize functionally correct code or to capture semantic similarity between natural language and code. Furthermore, the evaluation metrics and templates optimized for vulnerability detection do not align with the requirements of other tasks. For instance, they may not adequately measure fluency or correctness in code generation. This narrow focus restricts insights into the broader applicability of the proposed method across various code LLM applications, despite providing depth in analyzing its utility for vulnerability detection.
\end{enumerate}

\section{Conclusion}
This paper introduces CGP-Tuning, a code graph-enhanced, structure-aware soft prompt tuning method designed to improve code LLMs' understanding of code structures for better vulnerability detection. Experimental results demonstrate the importance of incorporating graph-based structural information for boosting the vulnerability detection performance of code LLMs. Unlike existing graph-enhanced soft prompt tuning methods, CGP-Tuning's cross-modal alignment module effectively captures graph-text interactions while maintaining computational efficiency. Moreover, when compared to various state-of-the-art methods, CGP-Tuning achieves superior performance and remains effective in detecting vulnerabilities within long source code. It is hoped that CGP-Tuning can offer the community a better alternative for fine-tuning code LLMs, enabling more accurate vulnerability detection by integrating structural and semantic information within code graphs, and ultimately contributing to more robust and intelligent security analysis techniques in software development.

\bibliographystyle{IEEEtran}
\bibliography{references}

%
\begin{IEEEbiography}[{\includegraphics[width=1in,height=1.25in,clip,keepaspectratio]{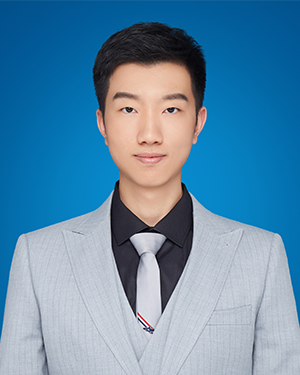}}]{Ruijun Feng}
is currently working toward a Ph.D. in Computer Science and Engineering at the University of New South Wales (UNSW), Sydney. His research interests include artificial intelligence for software engineering, focusing on combining large language models with security techniques to enhance code analysis, vulnerability detection, and secure software development practices.
\end{IEEEbiography}

\begin{IEEEbiography}[{\includegraphics[width=1in,height=1.25in,clip,keepaspectratio]{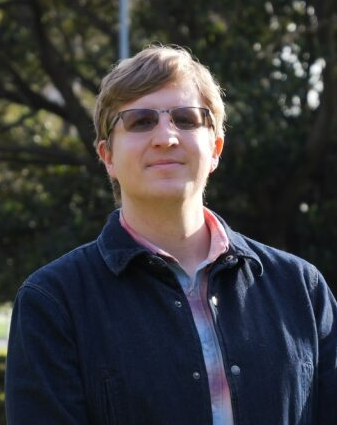}}]{Hammond Pearce}
(Member, IEEE) received the B.E. (Hons.) and Ph.D. degrees in computer systems engineering from The University of Auckland, Auckland, New Zealand. He is currently a Senior Lecturer (equivalent to an Assistant Professor) with the School of Computer Science and Engineering, University of New South Wales, Sydney, NSW, Australia. Previously, he was a Research Assistant Professor with the Department of Electrical and Computer Engineering, New York University, Brooklyn, NY, USA, and the NYU Center for Cybersecurity. His main research interests include large language models and cybersecurity, with a particular focus on hardware and embedded systems applications.
\end{IEEEbiography}

\begin{IEEEbiography}[{\includegraphics[width=1in,height=1.25in,clip,keepaspectratio]{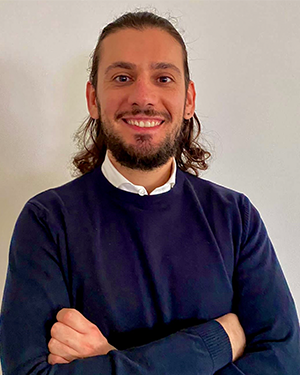}}]{Pietro Liguori}
is an Assistant Professor with the Department of Electrical Engineering and Information Technology at the University of Naples Federico II, Italy. His research interests span software security, generative AI, and automated vulnerability analysis. His recent work focuses on enhancing the robustness of AI-based code generators and developing static and dynamic techniques for assessing AI-generated software artifacts. He received a Ph.D. in Information Technology and Electrical Engineering from the University of Naples Federico II.
\end{IEEEbiography}

\begin{IEEEbiography}[{\includegraphics[width=1in,height=1.25in,clip,keepaspectratio]{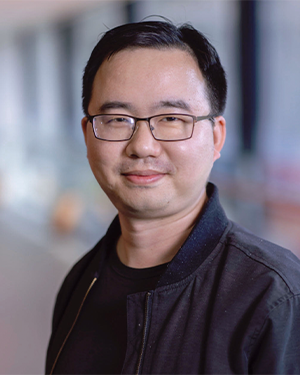}}]{Yulei Sui}
(Senior Member, IEEE) is an associate professor with the School of Computer Science and Engineering, University of New South Wales (UNSW). He is broadly interested in program analysis, secure software engineering and machine learning. In particular, his research focuses on building open-source frameworks for static analysis and verification techniques to improve the reliability and security of modern software systems.
\end{IEEEbiography}

\end{document}